\documentclass[useAMS,usenatbib,usegraphicx]{mn2e}

\title[Weak Lensing by Triaxial Dark Matter Halos]{A Statistical Study of Weak Lensing by Triaxial Dark Matter Halos: Consequences for Parameter Estimation}
\author[V. L. Corless \& L. J. King]{Virginia L. Corless$^{1}$\thanks{E-mail:
vc258@ast.cam.ac.uk} and Lindsay J. King$^{1}$\\
$^{1}$Institute of Astronomy, University of Cambridge, Madingley Road, Cambridge, United Kingdom}

\begin{document}

\date{2006 November 29}

\pagerange{\pageref{firstpage}--\pageref{lastpage}} \pubyear{2006}

\maketitle

\label{firstpage}

\begin{abstract}
Measuring the distribution of mass on galaxy cluster scales is a crucial test of the $\Lambda$CDM model, providing constraints on the nature of dark matter.  Recent work investigating mass distributions of individual galaxy clusters using gravitational lensing has illuminated potential inconsistencies between the predictions of structure formation models relating halo mass to concentration and those relationships as measured in massive clusters.  However, such analyses typically employ only simple spherical halo models with canonical NFW slopes, while the halos formed in simulations show a range of more complex features.   Here we investigate the impact of such expected deviations from the canonical NFW halo profile on mass and parameter estimation using weak gravitational lensing on massive cluster scales. Changes from the canonical NFW profile slopes are found to affect parameter estimation.  However, the most important deviation is halo triaxiality because it is impossible even with fiducial weak lensing data to fully resolve the three-dimensional structure of the halo due to lensing's sensitivity only to projected mass.  Significant elongation of the halo along the line of sight can cause the mass and concentration to be overestimated by as much as $50\%$ and by a factor of 2, respectively, while foreshortening has the opposite effect.  Additionally, triaxial halos in certain orientations are much better lenses than their spherical counterparts of the same mass, indicating that clusters chosen for study because of evident lensing are likely to be drawn from the high-triaxiality end of the halo shape distribution; cluster samples chosen with no shear bias return correct average parameter values.  While the effects of triaxiality alone may not be enough to fully explain the very high concentrations reported for some clusters, such as Abell 1689, they go a long way in easing the tensions between observations and the predictions of the cold dark matter paradigm.   
\end{abstract}

\begin{keywords}
gravitational lensing - cosmology: theory - dark matter - galaxies:clusters: general.
\end{keywords}
\section{Introduction}
Galaxy clusters are ideal laboratories in which to study dark matter, being the most massive bound structures in the universe and dominated by their dark matter component ($\sim90\%$). Constraining the clustering properties of dark matter is crucial for refining structure formation models that predict both the shapes of dark matter halos and their mass function (e.g. \cite{n1}; \cite{b1}; \cite{d2}).  Several methods are used to measure galaxy cluster dark matter profile shapes and halo masses on a range of scales, including X-ray studies, dynamical analyses, Sunyaev-Zeldovich surveys, and gravitational lensing.  However, all of these methods require simplifying assumptions to be made regarding the shape and/or dynamical state of the cluster in order to derive meaningful constraints from available data.  Most parametric methods typically assume spherical symmetry of the halo and X-ray and dynamical estimates additionally assume virialization of the cluster.  However, examination of halos in CDM structure formation simulations (e.g. \cite{b2} (using the Millennium simulation); \cite{s3}) and observed galaxy clusters show both of these assumptions to be unphysical; simulations show significant triaxiality in cluster-scale halos and observed galaxy clusters often exhibit complex dynamics that suggest recent or ongoing mergers and disruption (e.g. Cl0024+1654, \cite{c2};  1E0657-558, \cite{c3}).  Understanding the impact of these physical realities on cluster mass and parameter estimates is crucial for accurate comparisons between measured cluster properties and model predictions.

We focus on the impact of deviations from three assumptions frequently made in weak lensing analyses of galaxy clusters.  Gravitational lensing is an appealing tool for cluster studies because it is sensitive only to the projected mass and thus requires no assumptions be made about the dynamical state of the cluster.  However, because the signal-to-noise ratio is low for weak lensing measurements and there is significant degeneracy along the line-of-sight, only very simple parametric models can be fit. Even using space-based weak lensing data (where the number density of background sources useful for a weak lensing analysis is a factor of a few higher than that of ground-based data), it is often difficult to distinguish between a singular isothermal sphere (SIS) with mass density as a function of distance from halo centre $r$ given by $\rho(r) \propto r^{-2}$ and the universal dark matter profile of \cite{n1} (NFW) with $\rho(r) \propto r^{-1}$ in the innermost regions and $\rho(r) \propto r^{-3}$ in the outer regions.

Most cluster profile fits are carried out in the hope of either supporting or refuting the universality of the NFW profile and thus testing the CDM paradigm.  The NFW is typically parameterized by an approximate virial mass $M_{200}$ and a concentration parameter, $C$, and simulations predict a strong correlation between the two.  For a cluster of $M=10^{15}$ M$_{\odot}$, $C \sim 4$.  However, several authors (e.g. \cite{l2}; \cite{k5}; \cite{g2}) have recently reported results in the very low probability tail of the predicted distribution; notably, in a combined weak and strong lensing analysis of Abell 1689, \cite{b3} report a concentration parameter of $C=14 \pm 1.5$, when $C\sim4$ is expected.  While one such result is not damning, especially given the very complex, likely not relaxed, structure of A1689 described recently by \cite{l1}, it is nonetheless of interest to investigate how possible future discrepancies between observations and the predictions of $\Lambda$CDM should be interpreted.  

Crucially, more advanced N-body simulations carried out since the ground-breaking work of NFW indicate that cluster-scale dark matter halos are expected to be significantly triaxial, with axis ratios between minor and major axes as small as 0.4 (\cite{s3}).  \cite{o2} applied a fully triaxial NFW model to the shear map of Abell 1689 to find that it is consistent with $6\%$ of cluster-scale halos, and \cite{g1} showed that a triaxial NFW can reconcile parameter values derived from observations of the cluster MS2137-23 to predictions from N-body simulations. Here we carry out a general study of the statistical effects of triaxiality on weak lensing analysis.  In addition to the question of triaxiality, there is also significant scatter in the inner and outer slopes of NFW-like profiles in simulations.  Even the mean value of the inner slope is contentious, with some groups (e.g. \cite{d1}; \cite{t1}) finding it to be steeper than NFW. These effects have been explored in various combinations in the context of strong lensing, for example \cite{m5} investigated the interplay of triaxiality and slope constraints in strong lensing analyses.

We seek to measure the effects of these three alterations -- triaxiality and variations in the inner and outer slopes -- to the spherical NFW profile on weak lensing parameter and mass estimation.  Given the difficulty of distinguishing even entirely different classes of potential dark matter profiles using weak lensing data it is not currently feasible to meaningfully constrain complex NFW-like models with larger numbers of parameters that allow for such variations. 
There are several useful elliptical NFW models obtained by perturbing the lensing potential of a spherical NFW model; however, these are only realistic for low ellipticities and do not account for full triaxial structure (e.g. \cite{g3}, \cite{m6}).  We therefore estimate the effects of the three expected deviations from a spherical NFW on the model parameters and cluster masses derived fitting the simplest models (SIS, standard NFW and the isothermal ellipsoid SIE) to weak lensing data. Here we are interested in the properties of individual clusters; stacking the lensing signal from a sample of clusters gives a much higher signal-to-noise for the determination of their average profile (e.g. \cite{m3}).  Also note that we consider here the impact of cluster morphology rather than structure along the line of sight (for more on line of sight structure, see e.g. \cite{h1}).
In the next section we introduce the three altered NFW-like profiles and describe their lensing properties as implemented in the simulations described in section 3.  We present our results in section 4, and discuss our findings in section 5.

\section{Lensing by NFW-like Halos}
\subsection{Weak Lensing Background}\label{subsec:wl}
Weak lensing distorts the shapes and number densities of background galaxies.  The shape and orientation of a background galaxy can be described by a complex ellipticity $\epsilon^s$, with modulus $|\epsilon^s|=(1-b/a)/(1+b/a)$, where $b/a$ is the minor:major axis ratio, and a phase that is twice the position angle $\phi$, $\epsilon^s=|\epsilon^s|e^{2i\phi}$.  The galaxy's shape is distorted by the weak lensing reduced shear, $g=\gamma/(1-\kappa)$, where $\gamma$ is the lensing shear and $\kappa$ the convergence, such that the ellipticity of the lensed galaxy $\epsilon$ becomes
\begin{equation}\epsilon = \frac{\epsilon^s + g}{1 + g^{\ast}\epsilon^s} \approx \epsilon^s + \gamma\label{eq:lens}\end{equation}
in the limit of weak deflections.  The distributions of ellipticities for the lensed and unlensed populations are related by 
\begin{equation}p_{\epsilon} = p_{\epsilon^s}\left|\frac{d^2\epsilon^s}{d^2\epsilon}\right|;\end{equation}
assuming a zero-mean unlensed population, the expectation values for the lensed ellipticity on a piece of sky is $<\epsilon> = g \approx \gamma$.  This is the basis for weak lensing analysis in which the shapes of images are measured to estimate the shear profile generated by an astronomical lens.

\begin{figure*}
 \includegraphics[width=170mm, height=53.3mm]{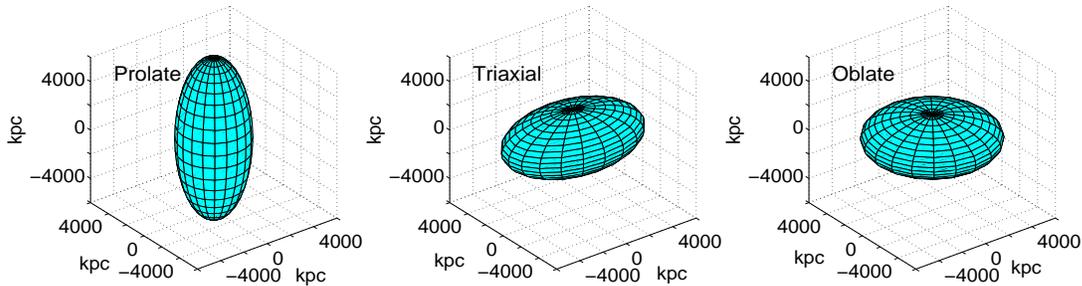}
 \caption{Prolate (a=b=0.4), triaxial (a=0.42, b=0.65), and oblate (a=0.4, b=1.0) clusters of equal mass.}
 \label{fig:ClusterPics}
\end{figure*}

Lensing also changes the number counts of galaxies on the sky via competing effects; some faint sources in highly magnified regions are made brighter and pushed above the flux limit of the observation, but those same regions are stretched by the lensing across a larger patch of sky and so the number density of sources is reduced.  Thus the number of sources in the lensed sky $n$ is related to that in the unlensed background $n_0$ and the slope of the number counts of sources at a given flux limit $\alpha$ by $n=n_0\mu^{\alpha-1}$, where $\mu$ is the lensing magnification $\mu^{-1}=(1-\kappa)^2 - |\gamma|^2$.  A full description of these effects is given in \cite{c1}.

We now describe the characteristic behaviour of the convergence $\kappa$ and shear $\gamma$ of the three analytic NFW-like density profiles we study.

\subsection{Triaxial NFW}
A full parameterization for a triaxial NFW halo is given by \cite{j1} (hereon JS02).  They generalize the spherical NFW profile to obtain a density profile
\begin{equation}
\rho(R) = \frac{\delta_c \rho_c(z)}{R/R_s(1 + R/R_s)^2}
\label{eq:3axrho}
\end{equation}
where $\delta_c$ is the characteristic overdensity of the halo, $\rho_c$ the critical density of the Universe at the redshift $z$ of the cluster, $R_s$ a scale radius, $R$ a triaxial radius 
\begin{equation}
R^2 = \frac{X^2}{a^2} + \frac{Y^2}{b^2} + \frac{Z^2}{c^2},\textrm{         }(a\leq b \leq c = 1),\label{eq:3axR}\end{equation}
and $a/c$ and $b/c$ the minor:major and intermediate:major axis ratios, respectively.  In a  different choice from JS02 we define a triaxial virial radius $R_{200}$ such that the mean density contained within an ellipsoid of semi-major axis $R_{200}$ is $200\rho_c$ such that the concentration is
\begin{equation}C = \frac{R_{200}}{R_s},\label{eq:3axC}\end{equation}
the characteristic overdensity is
\begin{equation}\delta_c = \frac{200}{3} \frac{C^3}{\log (1+C) - \frac{C}{1 + C}},\label{eq:3axdelta}\end{equation}
the same as for a spherical NFW profile, and the virial mass is
\begin{equation}M_{200} = \frac{800\pi}{3}abR_{200}^3\rho_c.\label{eq:3axM200}\end{equation}
This differs significantly from the parameterization of JS02 in which the ellipsoidal virial radius is defined in terms of an overdensity dependent on the axis ratios.  Further, there the virial mass is defined in terms of an effective spherical virial radius that is a constant fraction of the ellipsoidal virial radius, making it independent of the axis ratios of the ellipsoid.

While the parameterization of Jing $\&$ Suto has the appealing property of giving the virial mass as a function only of the virial radius with no need for reference to the axis ratios of the triaxial halo, it is not ideal for our purposes.  Our choice of parameters treats the ellipsoid as such all along without approximation and gives an effective concentration of $R_{200}/R_s$ along each of the three halo axes.  Further, our choice of an overdensity at collapse independent of axis ratio is well motivated by ellipsoidal collapse models that predict collapse to stop at the same enclosed density as does spherical collapse (\cite{s2}).  Additionally, the advantages of deriving a mass from knowledge of the virial radius alone are mostly lost in the context of profile fitting to observational data, as the axis ratios of the cluster enter into the characteristic overdensity and thus are necessary parameters in any fit of the triaxial model.  Most crucial to any work is consistency when comparing parameters across different bodies of work; to this end we include in Appendix A conversions and comparisons between our parameterization and that of JS02, as well as to parameters derived by fitting to a spherically averaged density profile, similar to those often quoted for N-body simulations.

Though all our formalism is valid for a general triaxial halo, we focus primarily on symmetric prolate and oblate halos, the first with two equal-length short axes and the second with two equal-length long axes.  For ease of distinguishing between these representative halos we introduce an axis ratio $Q$, defined as the ratio between the odd axis and the similar axes, so that a prolate halo with $a=b=0.3$ and $c=1$ has an axis ratio $Q=3.33$, while an oblate halo with $a=0.3$ and $b=c=1$ has an axis ratio $Q=0.3$.  Moving from small to large axis ratios $Q$ is thus equivalent to beginning with a very flat, oblate halo (a ``pancake''), stretching it along the short axis until it becomes spherical ($Q=1$), and continuing to stretch the same axis past spherical, finally ending with a very prolate, ``cigar''-like halo.

\subsubsection{Lensing Properties}
The full derivation of the lensing properties of a triaxial halo is given by \cite{o1} (hereon OLS), and we summarize some of that work here.\footnote{Although we set $c=1$, we keep $c$ as a variable in our notation for consistency with OLS.}  The triaxial halo is projected onto the plane of the sky to find its projected elliptical isodensity contours as a function of the halo's axis ratios and orientation angles ($\theta$, $\phi$) with respect to the the observer's line-of-sight.  The elliptical radius is given by
\begin{equation}\zeta^2 = \frac{X^2}{q_X^2} + \frac{Y^2}{q_Y^2}\label{eq:xi}\end{equation}
where $(X,Y)$ are physical coordinates on the sky with respect to the centre of the halo,
\begin{eqnarray}
q_X^2&=&\frac{2f}{\mathcal{A}+\mathcal{C} - \sqrt{(\mathcal{A}-\mathcal{C})^2 + \mathcal{B}^2}}\\
q_Y^2&=&\frac{2f}{\mathcal{A}+\mathcal{C} + \sqrt{(\mathcal{A}-\mathcal{C})^2 + \mathcal{B}^2}} \end{eqnarray}
where
\begin{equation}f = \sin^2\theta\left(\frac{c^2}{a^2}\cos^2\phi + \frac{c^2}{b^2}\sin^2\phi\right) + \cos^2\theta,\label{eq:f}\end{equation}
and
\begin{eqnarray}
\mathcal{A}&=&\cos^2\theta\left(\frac{c^2}{a^2}\sin^2\phi + \frac{c^2}{b^2}\cos^2\phi\right) + \frac{c^2}{a^2}\frac{c^2}{b^2}\sin^2\theta,\\
\mathcal{B}&=&\cos\theta\sin 2\phi\left(\frac{c^2}{a^2} - \frac{c^2}{b^2}\right),\\
\mathcal{C}&=&\frac{c^2}{b^2}\sin^2\phi + \frac{c^2}{a^2}\cos^2\phi.
\end{eqnarray}
The axis ratio $q$ of the elliptical contours is then given by
\begin{equation}q = \frac{q_Y}{q_X}\label{eq:q}\end{equation}
and their orientation angle $\Psi$ on the sky by
\begin{equation}\Psi = \frac{1}{2}\tan ^{-1}\frac{\mathcal{B}}{\mathcal{A}-\mathcal{C}}~~~(q_X \ge q_Y).\label{eq:psi}\end{equation}

Here we diverge slightly from OLS's treatment as we are interested not in deflection angles but in the lensing shear and convergence, both combinations of second derivatives of the lensing potential $\Phi$ (commas indicate differentiation):
\begin{eqnarray}
\gamma_1&=& \frac{1}{2}\left(\Phi_{,XX} - \Phi_{,YY}\right),\\
\gamma_2 &=&\Phi_{,XY},\\
\kappa &= &\frac{1}{2}\left(\Phi_{,XX} + \Phi_{,YY}\right).
\end{eqnarray}
These derivatives are calculated as functions of integrals of the spherical convergence $\kappa(\zeta)$ (see e.g. \cite{b4} for a full treatment of weak lensing by a spherical NFW profile) following the method of \cite{s5} and \cite{k1}, normalized by a factor of $1/\sqrt{f}$ from Equation~\ref{eq:f} (see OLS for the derivation of this normalization)
\begin{eqnarray}
\Phi_{,XX} &= &2qX^2K_0 + qJ_0,\\
\Phi_{,YY} &= &2qY^2K_2 + qJ_1,\\
\Phi_{,XY} &= &2qXYK_1,\end{eqnarray}
where
\begin{eqnarray}K_n(X,Y)& =&\frac{1}{\sqrt{f}}\int_0^1 \frac{u\kappa'(\zeta(u)^2)}{[1 - (1-q^2)u]^{n+1/2}}du,\label{eq:K}\\
J_n(X,Y)& =&\frac{1}{\sqrt{f}}\int_0^1 \frac{\kappa(\zeta(u)^2)}{[1 - (1-q^2)u]^{n+1/2}}du,
\end{eqnarray}
and
\begin{equation}\zeta(u)^2 = \frac{u}{q_X}\left(X^2 + \frac{Y^2}{1 - (1-q^2)u}\right).\end{equation}

Note that our radial variable $\zeta$ appears different from Keeton's $\xi$ because it is defined in terms of two axis ratios $q_X$ and $q_Y$ rather than one $q$: $\zeta = \xi/q_X$.  This reflects a dependence on the 3D structure of the cluster; for example, extended structure along the line of sight decreases $q_X$ and thus increases the convergence and shear at a given $(X,Y)$.

\begin{figure}
 \includegraphics[width=80mm, height=120mm]{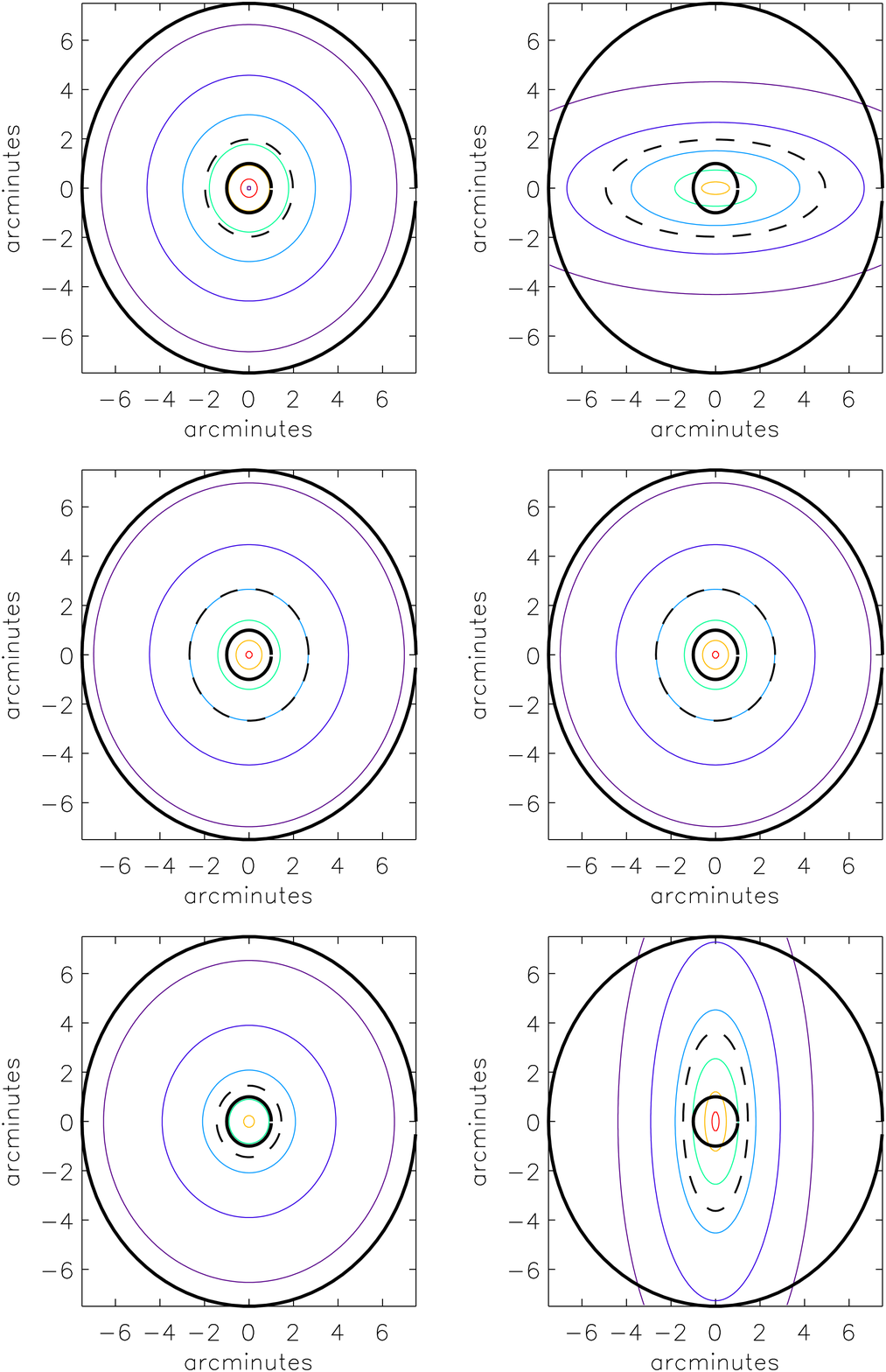}
 \caption{The top (middle) \{lower\} panels show  isoconvergence contours for a prolate (spherical) \{oblate\} halo of $M_{200}=10^{15}$ M$_{\odot}$, $C=4$, with axis ratios $a=b=0.4$ ($a=b=1.0$) \{$a=0.4, b=1.0$\}; the left-hand panel shows the halo oriented with the odd axis along the line of sight, the right-hand panel shows the halo with odd axis in the plane of the sky.  The thick solid lines show the limits of the aperture from which weak lensing data is taken, and the dashed line shows the $R_s$ ellipse for each projection (note that $R_s$ is scaled by the axis ratio in each direction, so that when looking at the minor axis of a triaxial halo, the apparent scale radius is $a$ times the $R_s$ value of the halo).  The lowest contour corresponds to $\kappa = 0.02$ and subsequent contours each increase in $\kappa$ by a factor of 2.}
 \label{fig:project}
\end{figure}

Figure~\ref{fig:project} shows the convergence $\kappa$ for a symmetric prolate (``cigar''-shaped), spherical, and oblate (``pancake''-shaped) halo, each of mass $M_{200} = 10^{15}$ M$_{\odot}$.  On the left the halos are oriented so that the odd axis is along the line of sight --  the long axis of a prolate halo and the short axis for an oblate halo -- on the right they are oriented so that the odd axis is completely in the plane of the sky.  The prolate halo oriented with odd axis along the line of sight has the highest convergence due to the large quantity of projected mass hidden in it, while the oblate halo in the same configuration has very low values of $\kappa$.  When viewed with the odd axis in the plane of the sky the situation is reversed; the prolate halo has low convergence due to a smaller amount of mass along the line of sight while the oblate halo has higher convergence values.

Viewing halos in projection is sometimes counterintuitive; for halos of equal triaxial virial mass the 3-D major axis of a prolate halo is longer than that of an oblate halo, which is in turn longer than the 3-D radius of a spherical halo.  This is not apparent in projection, where, when the odd axis is in the plane of sky, the larger total amount of mass in the halo along the line of sight makes the isodensity contours of the oblate halo larger than those of its prolate counterpart.

\subsection{Varying slopes}
We consider profiles in which only one of the inner or outer slopes is allowed to vary from its canonical NFW value.

\subsubsection{Steeper inner slope}
Several groups (\cite{j2}, \cite{k2}, \cite{w1}) have studied a generalized NFW with a fixed outer slope $\rho \propto r^{-3}$ and a varying inner logarithmic slope $-m$ with density profile
\begin{equation}\rho(r) = \frac{\delta_c \rho_c}{(r/r_s)^{m}  \left(1+ r/r_s\right)^{3-m}}.\label{eq:freerho}\end{equation}
Its lensing properties cannot be computed analytically; the convergence is written most simply as
\begin{eqnarray}
\kappa(r) &=& \frac{2\delta_c\rho_c r_s}{\Sigma_{cr}} x^{1-m}\bigg[(1+x)^{m-3}+ \nonumber\\
&&\left.(3-m)\int_0^1(y+x)^{m-4}\left(1-\sqrt{1 - y^2}\right)dy\right],\label{eq:freekappa}\end{eqnarray}
where $x=r/r_s$ and $\Sigma_{cr}$ is the lensing critical surface density.  The shear is calculated according to the relationship $|\gamma| = \bar{\kappa} - \kappa$ (\cite{m4}), true for any spherically symmetric lens, in which $\bar{\kappa}$ is the dimensionless mean enclosed surface density and is given by
\begin{eqnarray}\bar{\kappa}(r)& = &\frac{4\delta_c\rho_c r_s}{\Sigma_{cr}}x^{1-m}\times\nonumber\\
&&\bigg\{\frac{1}{3-m}\,_2F_1\left[3-m, 3-m; 4-m : -x\right] +\nonumber\\
&& \left.\int_0^1(y+x)^{m-3}\frac{1-\sqrt{1-y^2}}{y}dy\right\},\label{eq:freekbar}\end{eqnarray}
where $_2F_1$ is the hypergeometric function.  For a spherically symmetric lens, the shear is always purely in the tangential direction.  The characteristic overdensity $\delta_c$ is given by
\begin{equation}\delta_c =\frac{200(3 - m)C^{m}}{3 _2F_1[3-m, 3-m; 4- m: -C]},\label{eq:freedeltac}\end{equation}
where $C=r_{200}/r_s$ is the concentration parameter and $r_{200}$ is the virial radius~\citep{k2}.\footnote{Keeton $\&$ Madau use a different definition of $C_{-2} = r_{200}/r_{-2}$ where $r_{-2}$ is the radius at which the logarithmic slope of the density is $-2$.  They give a simple transformation between the two definitions, $C = (2-m)C_{-2}$.}  Note that this expression for the overdensity differs by one power of $C$ from Equation 2 in Keeton $\&$ Madau, correcting a typo in the original paper.  The virial mass is simply $M_{200} = \frac{4}{3}\pi 200\rho_c r_{200}^3$.
\subsubsection{Varying outer slope}
The density profile of an NFW-like halo with an outer slope left free to vary can be written
\begin{equation}\rho(r) = \frac{\delta_c \rho_c}{(r/r_s)(1 + r/r_s)^{n-1}}\label{eq:ofreerho}\end{equation}
where the exponent $n-1$ is chosen so that $n=3$ corresponds to the canonical NFW logarithmic outer slope of -3.
The characteristic overdensity is then given by 
\begin{equation}\delta_c = \frac{200}{3}\frac{C^3(n-2)(n-3)}{1 - [1 + (n-2)C](1+C)^{2-n}},\label{eq:ofreedelta}\end{equation}
where $C=r_{200}/r_s$; there is no simple expression for the convergence or shear so they must be calculated numerically by integrating through the mass distribution along the line of sight and employing the relationship between $\bar{\kappa}$ and $\gamma$:
\begin{eqnarray} \kappa(r) &=& \frac{1}{\Sigma_{cr}}\int_{-\infty}^{\infty}\rho(\sqrt{r^2 + z^2})dz\\
\bar{\kappa}(r) &=& \frac{2}{r^2}\int_{0}^r r'\kappa(r') dr'\\
|\gamma|&=&\bar{\kappa} - \kappa.
\end{eqnarray}

\subsection{Parametric Fits}
Three simple parametric models are fit to the NFW-like models described above.  They are the spherical NFW, the Singular Isothermal Sphere (SIS) and the Singular Isothermal Ellipsoid (SIE) (a good summary of their lensing behavior is found in \cite{k4}).  They have respectively 2, 1, and 3 free parameters and are the models most often fit to lensing clusters.  Each family is characterized by a set of parameters $\Pi$: \{$C$, $M_{200}$\}, \{Einstein radius $\theta_{E}$\}, and \{$\theta_E$, axis ratio $q$, orientation angle $\Psi$\} respectively.  The best-fit parameters are obtained from a given lensed catalogue of $n_{\gamma}$ galaxies, each with ellipticity $\epsilon_i$ and position $\vec{\mathcal{\theta}_i}$ by minimizing the shear log-likelihood function (\cite{s1}, \cite{k4})
\begin{equation}\ell_{\gamma} = -\sum_{i=1}^{n_{\gamma}}\ln  p_{\epsilon}(\epsilon_i|g(\vec{ \mathcal{\theta}_i}; \Pi)).\end{equation}

\begin{figure}
 \includegraphics[width=80mm, height=75mm]{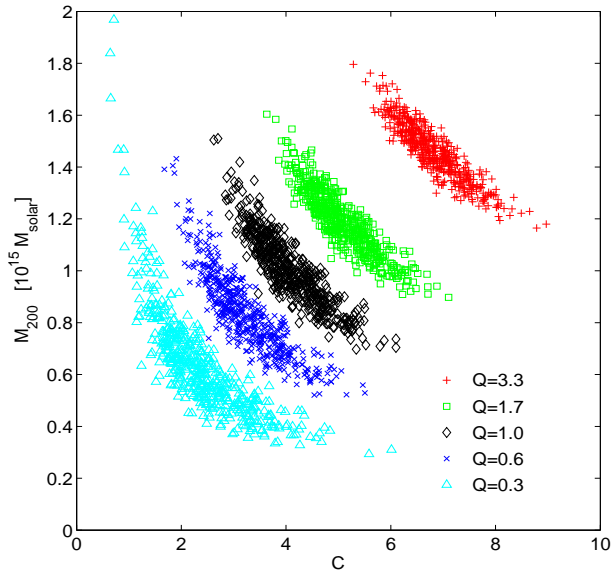}
 \caption{Best-fit concentration $C$ and virial mass $M_{200}$ for a simple NFW model fit to five triaxial lenses with $C=4$, $M_{200} = 10^{15}$ M$_{\odot}$, circularly symmetric on the sky with odd axis oriented along the line of sight.  Halos with axis ratio $Q$ (ratio between odd axis and similar axes) less than one are oblate ($Q=0.3 \rightarrow a=0.3$), while those with $Q$ greater than one are prolate ($Q=3.3 \rightarrow a=b=0.3$).}
 \label{fig:3axmultiplot}
\end{figure}

\section{Weak Lensing Simulations}

The main body of simulations is carried out for a field $7.5'$ in radius, with a background source density $n_0=30/$arcminute$^2$, typical of ground-based observations.  Although wide-field imaging of clusters with fields of $\sim$ 0.5 degrees is routinely possible, there are arguments that errors due to large-scale structure along the line-of-sight become more important as the shear due to the cluster itself diminishes with distance from the centre (e.g. \cite{h1}).  In any case, the general trends will hold for larger fields.  Poisson noise is accounted for.  A catalogue of randomly positioned and oriented galaxies with intrinsic shapes $\epsilon^s$ drawn from a Gaussian distribution with dispersion $\sigma=0.2$ in the modulus $|\epsilon^s|$ is placed at redshift $z=1$.  This catalogue of background galaxies is lensed through a model lens of choice placed at redshift $z=0.18$ (the redshift of Abell 1689), at which the width of the field is $\sim 1900$ kpc/h.  Thus our choice to place all sources on a sheet at $z=1$ is justified by the low redshift of our fiducial lens; only for higher redshift lenses that are in the heart of the redshift distribution is the distribution of source redshifts important \citep{s4}.  The background galaxies are lensed according to Equation \ref{eq:lens} and the number counts are reduced as prescribed in section~\ref{subsec:wl}, taking the slope of the source number counts in flux to be ${\rm d log}N/{\rm d log}S=\alpha=0.5$ (corresponding to a slope of 0.2 in magnitude as in \cite{f1}).  Galaxies located within $1'$ of the cluster centre are removed from the analysis to avoid the strong lensing regime at the centre of the cluster (in any case background galaxies near the cluster centre would be mostly obscured by cluster members in observations).  Throughout we assume a concordance cosmology with $\Omega_m=0.3$, $H_0 = 70$ km s$^{-1}$ Mpc$^{-1}$, $h=0.7$, and a cosmological constant $\Omega_{\Lambda} = 0.7$, and a typical massive cluster of triaxial $M_{200}=10^{15}$ M$_{\odot}$ and $C=4$, with corresponding scale radii of $R_s = $\{340.38 kpc/$h$, $461.97$ kpc/$h$, 626.99 kpc/$h$\} for the spherical, oblate ($a=0.4, b=1$), and prolate ($a=b=0.4$) cases.  At the redshift of our lens $z=0.18$, 1 arcminute corresponds to $\sim 127$ kpc/$h$.

\begin{figure}
 \includegraphics[width=80mm, height=80mm]{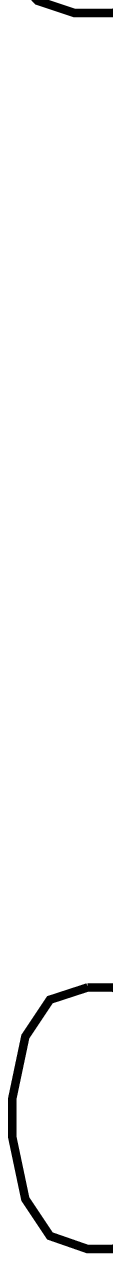}
 \caption{The top (bottom) panel shows the mean best-fit NFW mass $M_{200}$ (concentration $C$) after 500 lensing realizations as a function of axis ratio $Q$ for oblate and prolate triaxial halos oriented along the line of sight such that the observed density contours are circular, for halo concentrations $C=4$ and $C=8$.  For small $Q$ the halo is ``pancake'' shaped and for large $Q$ it is ``cigar'' shaped.  Error bars show the approximate 1$\sigma$ dispersion within each set of realizations.}
 \label{fig:M3axMeans}
\end{figure}

\section{Results}
\subsection{Fits to Lensing Data from Triaxial Halos}
To characterize the impact of triaxiality on parameter estimation over a range of halo shapes we choose ten representative triaxial halos to model; five symmetric oblate halos with $b=c=1$ and five symmetric prolate halos with $c=1$, $a=b$, and $a=\{0.3, 0.4, 0.6, 0.8, 0.9\}$ for both sets.  Each halo is studied in two orientations, one in which the odd-length axis is oriented along the line of sight (in which case an observer would see a circularly symmetric shear pattern on the sky), and one in which the odd length axis is oriented in the plane of the sky (in which case maximum ellipticity is observed).  These will be referred to as the LoS and Plane cases respectively.  500 simulations are carried out for every halo type and orientation, and NFW and SIE models are fit to each of the lensed catalogues.

We further repeat the simulations using halos of the same mass with a higher concentration $C=8$ to test how sensitive our results are to changes in the underlying cluster parameters.  Due to the factor of 2 decrease in $R_s$ implied by this increase in concentation, we expect the overall quality of the fits to decrease as, for some orientations and axis ratios, $R_s$ will fall within the 1' annulus removed to avoid the strong lensing regime.  Throughout, plotted error bars are approximate 1$\sigma$ dispersions; though we expect the parameter dispersion within a set of realizations to be non-Gaussian, we use the standard deviation $\sigma^2 = \left(\frac{1}{N-1}\right)\sum_{i=1}^N(x_i - \bar{x})^2$ as a rough indicator of the relative dispersions of fits to different halo models.

\subsubsection{Hidden triaxiality: LoS halos}\label{sssec:LoSHalos}
Figure~\ref{fig:3axmultiplot} shows the parameters obtained fitting NFW profiles to 500 lensing realizations through several of the triaxial halos (see \cite{o2} for earlier work regarding the effects of triaxiality on the $M_{200}-C$ relationship); in Figure~\ref{fig:M3axMeans} mean best-fit concentrations and virial masses are plotted as a function of axis ratio $Q$ for all halos oriented along the line of sight.

We find that LoS oblate clusters are measured to have lower masses than true, while prolate clusters are measured to have higher masses;  in the extreme case of $a=0.3$, the mass of an oblate halo is underestimated by $\sim40\%$ and that of a prolate halo overestimated by $\sim45\%$.  Concentrations are also significantly affected by hidden triaxiality; for extreme oblate halos the measured concentration $C$ is half that of the true value while for prolate halos it is double!   Implied in the concentration and mass fits is a decrease in scale radius $r_s$ with increasing prolateness, as expected since clusters of equal mass and longer odd axes  have shorter circular axes; i.e. when one stretches the odd axis, one also shrinks the similar axes.  Thus for halos oriented with the odd axis along the line of sight, the extent of the halo on the plane of the sky will be smallest for very prolate halos and largest for very oblate ones.

We find the same type and scale of effects when the true concentration is increased to $C=8$. We also find similar trends when fitting SIE models, shown in Figure~\ref{fig:SIEallLow}, in that $\theta_E \propto M_{200}$ increases with increasing $Q$: in all cases studied a long axis oriented along the line of sight significantly increases mass and concentration estimates, while a short axis so oriented reduces them by a similar amount.

\begin{figure}
 \includegraphics[width=80mm, height=80mm]{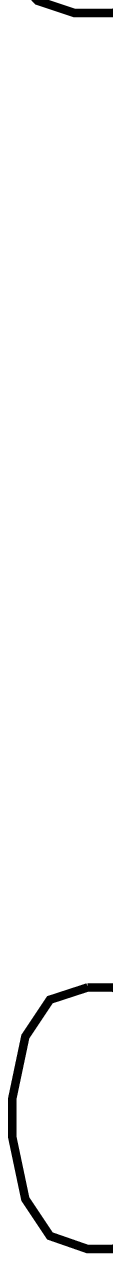}
 \caption{The top (bottom) panel shows the mean best-fit NFW mass $M_{200}$ (concentration $C$) after 500 realizations as a function of axis ratio $Q$ for oblate and prolate triaxial halos oriented with the odd axis in the plane of the sky such that the observed density contours are elliptical, for halo concentration $C=4$.  Error bars show the 1$\sigma$ dispersion within each set of realizations.}
 \label{fig:M3axSideMeans}
\end{figure}

\subsubsection{Visible triaxiality: halos in the plane}
Figure~\ref{fig:M3axSideMeans} plots the mean best-fit concentrations and virial masses as a function of axis ratio $Q$ for the studied triaxial halos oriented with the odd axis in the plane of sky.  Both prolate and oblate halos look elliptical on the sky in this orientation, as illustrated in Figure~\ref{fig:project}.  In this orientation we find that the mass is typically overestimated for oblate halos (in the most discrepant case of $Q=0.6$ by $\sim6\%$) and underestimated for prolate halos (for $Q=3.3$ by $\sim45\%$), as expected due to the greater mass hidden in projection in the oblate halos.  The degree of overestimation of the mass decreases slightly for the most extreme oblate axis ratios; this can be understood as the effect of increasing ellipticity on the sky decreasing the amount of mass included in the spherical halo fit to the elliptical distribution.  The concentration decreases with increasing $Q$, the opposite of the behaviour in the LoS case; however, the deviation form the true value is less in this case, with maximum $C\approx5$ and minimum $C\approx 3$.  

\begin{figure}
 \includegraphics[width=80mm, height=80mm]{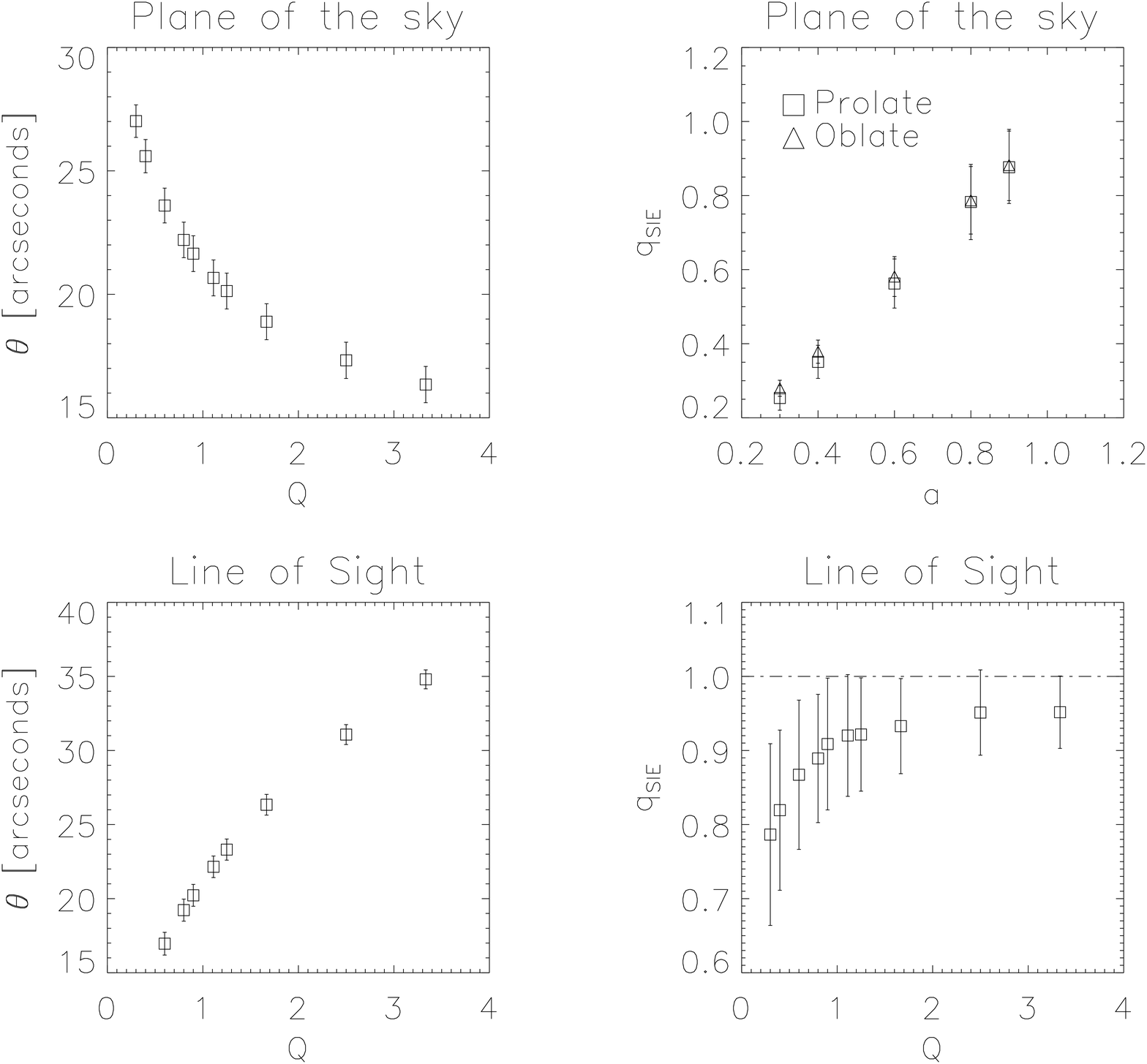}
 \caption{The best-fit SIE Einstein radius $\theta_E$ and visible axis ratio $q_{SIE}$ values for $C=4$ triaxial halos aligned in the plane of sky (along the line of sight) are plotted in the top (bottom) panels.  For the plane of the sky case, $q_{SIE}$ is plotted as a function of the projected triaxial axis ratio $q$ (defined for the triaxial NFW in Equation~\ref{eq:q}, where $q=a$, the minor axis ratio, for the plane of the sky case).  For the LoS cases, for which $q=1$, all values are plotted as a function of the odd:same axis ratio $Q$.  Error bars show the 1$\sigma$ dispersion within each set of realizations.  The $C=8$ halos exhibit similar behaviour, but give systematically higher values for $\theta_E$ (the maximum value of $\theta_E$ is found for the $q=0.3$ prolate LoS case, in which $\theta_E=46''$) and for the $q=1$ LoS cases give $q_{SIE}$ values closer to the true value than do their $C=4$ counterparts.}
 \label{fig:SIEallLow}
\end{figure}

When dealing with significant ellipticity visible on the sky it is of course hoped that observations will reveal it and better suited models can be fit, perhaps using higher quality data. We choose not to use the seemingly obvious choice of an approximate elliptical NFW for several reasons: such models are limited to relatively low axis ratios, cannot give a well-defined measure of the 3-D halo mass, and require four free parameters, often too many for a weak lensing analysis to constrain.  Instead, we fit an SIE profile to all ten halos; the best-fit SIE Einstein radius and axis ratios $q$ are plotted in Figure~\ref{fig:SIEallLow}.  $\theta_E$ and $M_{200}$ decrease with increasing $Q$, reproducing the general trend observed in the NFW fits.  The slight decrease in the best-fitting NFW mass at the lowest $Q$ values is not reproduced, further suggesting that the decrease observed in the NFW mass is an artifact of the specific behaviour of the profile under adverse fitting conditions.  The SIE generally does a good job of fitting the correct axis ratio, plotted in the figure against the actual axis ratio on the sky; the mean values of the axis ratios are systematically low by $\sim 0.05$ due to the imposition of a hard wall at $q=1$ in the minimization routine.  Because the offset  is constant across a large range of axis ratios the SIE emerges as an useful tool for detecting visible ellipticity, even when the lensing is a result of a non-isothermal distribution.  As a control we also fit the SIE to LoS halos, all of which have no apparent ellipticity on the sky, as shown in the bottom panel of Figure~\ref{fig:SIEallLow}.    We find the same systematic offset of $\sim0.05$ for the prolate halos; for oblate halos lower values of $q$ are often fit with large dispersions due to their low convergence and shear values in this orientation.  

\subsubsection{Likelihood contours for individual realizations}
\begin{figure}
 \includegraphics[width=80mm, height=80mm]{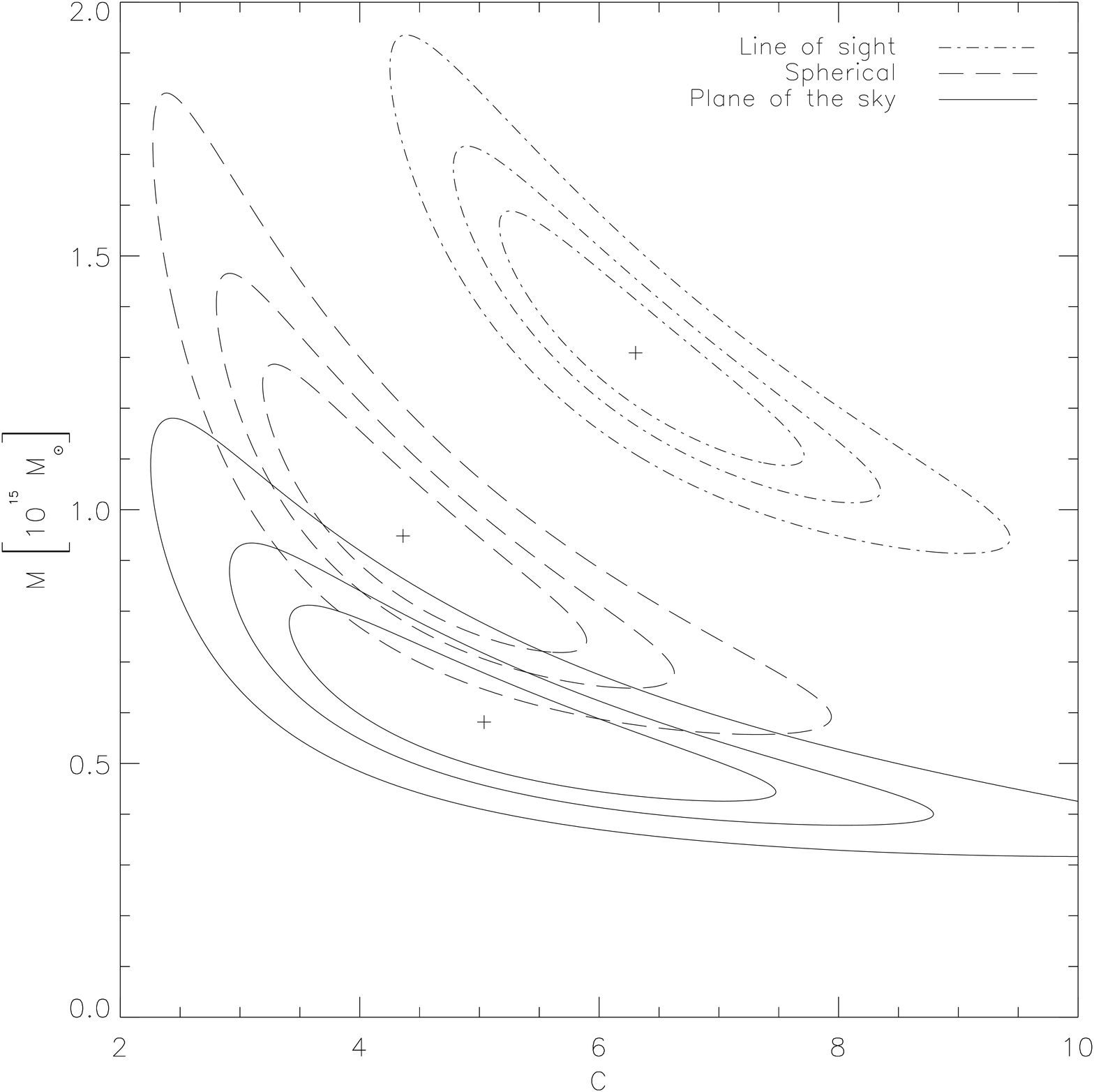}
 \caption{Contours of 70$\%$, 90$\%$, and 99$\%$ confidence for individual NFW fits to an $M_{200}=10^{15}$ M$_{\odot}$, $C=4$ prolate halo with $a=b=0.4$, $Q=2.5$, oriented along the line of sight and in the plane of the sky.  The contours for an NFW fit to a spherical halo of the same mass are shown for comparison.  Halos with $C=8$ exhibit contours of similar size and shape.}
 \label{fig:contours}
\end{figure}

Figure~\ref{fig:contours} plots isolikelihood contours obtained fitting an NFW to a single lensing realization through an oblate halo with $a=0.4$, $Q=2.5$, oriented in the plane of the sky and along the line of sight, as well as through a spherical halo of the same mass.  The contours are tightest for the LoS case; in this orientation the halo looks very much like a spherical NFW halo with higher mass than true, and so the lensing is stronger and the fit more constrained.  Conversely, the contours are larger than for the spherical case when the oblate halo is oriented in the plane of the sky, because in this case the spherical NFW is a poor fit to the elliptical isodensity contours of the lens.  The uncertainty of the fit is exacerbated by the lower convergence and shear generated by the halo in this configuration.

\begin{figure}
 \includegraphics[width=85mm, height=85mm]{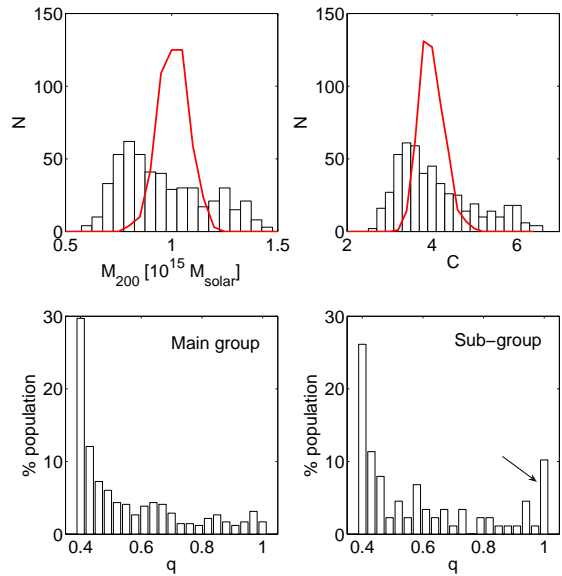}
 \caption{The top panels plot the distribution of NFW best-fit concentration and virial mass for a prolate halo with $a=b=0.4$ randomly oriented for 500 realizations.  The overplotted solid line shows the distribution for a spherical halo of the same mass.  The bottom panels show the distribution of apparent axis ratios on the sky of the lensing halos responsible for the two ``populations'' possibly apparent in the prolate halo parameter distributions.}
 \label{fig:orientplots}
\end{figure}
\subsubsection{Triaxial Halos Averaged Over Orientation}
In addition to the main body of simulations described above additional simulations were undertaken to determine the mean distortion caused in parameter and mass estimates for triaxial halos across all orientations.  For these a higher number density $n=120/$arcminute$^2$ was used, typical of space-based observations.  500 catalogues were lensed through the same halo, with the halo at a different random orientation for each catalogue.  This was done for three oblate halos and three prolate halos with $a=\{0.4, 0.8, 0.9\}$ for each class, and for one spherical model as a control.  NFW models were fit to each catalogue and the best-fit parameters averaged across the five hundred orientations.  We find that the mean parameter values remain very close to the true value, but that the dispersion around the mean increases with increasing triaxiality: the dispersion in the mass increases systematically from $\sigma_M=0.7\times10^{14}$ M$_{\odot}$ in the spherical case to $\sigma_M=1.0\times10^{14}$ M$_{\odot}$ for the most oblate halo and to $\sigma_M=2.0\times10^{14}$ M$_{\odot}$ for the most prolate halo, and similarly the dispersion in the concentration increases from $\sigma_C=0.3$ in the spherical case to  $\sigma_c=0.8$ for the most oblate halo and to $\sigma_C=0.9$ for the most prolate halo.  This trend is visible in Figure~\ref{fig:orientplots}, where the mass and concentration parameter distribution obtained for a spherical halo is overplotted on that obtained from an $a=b=0.4$, $Q=2.5$ prolate halo of the same mass.

\begin{figure}
 \includegraphics[width=80mm, height=80mm]{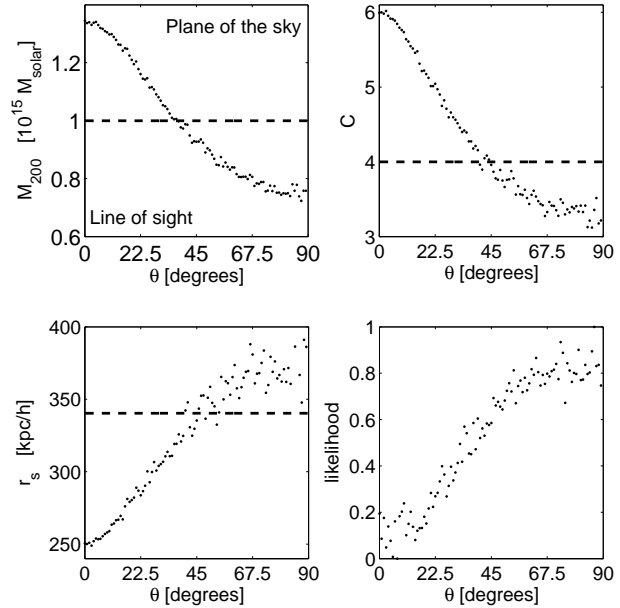}
 \caption{Best-fit NFW concentration, virial mass, and scale radius $r_s$ and the scaled likelihood of those fits for a prolate halo ($a=b=0.4$) rotated through $90^{\circ}$ on the sky. The dotted lines show the true parameter values for the lensing halo. The best-fits, highest concentrations, and highest masses are obtained when the long axis of the halo is oriented along the line of sight.}
\label{fig:plotorient}
\end{figure}

The distributions for the prolate halo are significantly more non-Gaussian than their spherical counterparts; even more, the distribution of $C$ values appears somewhat bimodal, with most halos falling in a Gaussian-like distribution around the true value $C=4$ but a significant number in a subpopulation centered at $C\approx 5.8$.  Plotting the axis ratios on the sky $q$ (as calculated in Equation~\ref{eq:q}) of the member halos of these two apparent ``populations,'' as is done in the bottom panels of Figure~\ref{fig:orientplots}, shows that the sub-group on the right contains most of the $q\approx1$, near LoS orientated lenses.  It seems then that halos with significant triaxiality visible in the plane of the sky are effectively radially averaged by fitting with a spherical model, and thus the parameters fit are evenly distributed near the true value, with a slight skew to low values.  However, halos that appear almost spherical on the sky are consistently fit as more massive, more concentrated halos in this case of an underlying prolate lens.  Oblate halos show the opposite behaviour, leading to consistent mass and concentration underestimates when LoS oriented.  This behaviour is illustrated for the $a=0.4$ prolate halo in Figure~\ref{fig:plotorient}, which plots parameters obtained from a single lensing realization under very low noise conditions in which the number density of sources $n_0$ is very high and the ellipticity dispersion $\sigma$ is very low, as well as a scaled likelihood ($0$ best, $1$ worst) as a function of orientation angle $\theta$.  As expected, the likelihood decreases significantly as the halo moves from LoS to Plane orientation due to the increasing poorness of fit of a spherical model to the increasingly elliptical profiles.

\subsubsection{Behavior of parameters as a function of $Q$}
\begin{figure}
 \includegraphics[width=80mm, height=80mm]{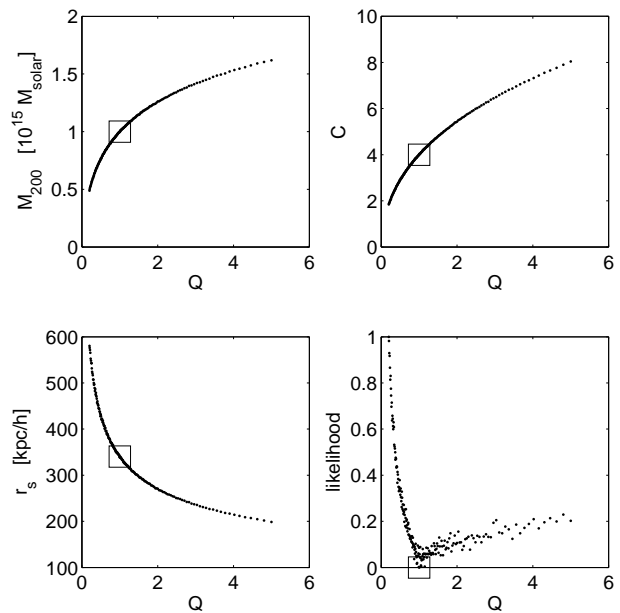}
 \caption{Best-fit NFW concentration, virial mass, and scale radius $r_s$ and the scaled likelihood of those fits for halos aligned along the line of sight as a function of axis ratio $Q$.  Boxes indicate the location of the spherical model.}
 \label{fig:plotplot}
\end{figure}

Catalogues simulated with very low ellipticity dispersion $\sigma$ and high number density $n_0$ minimize the effects of noise, generating, for Gaussian-like error distributions, best-fit parameters that represent the mean best-fit parameters that would be obtained after running many realizations for real noise conditions (\cite{s1}).  Catalogues were lensed under such reduced-noise conditions for many values of $a$ in both oblate and prolate triaxial models aligned with the odd-axis along the line of sight, to understand the ``worst-case'' behaviour of parameter estimates across a range of axis ratios.  Figure~\ref{fig:plotplot} shows that the mass and concentration estimates increase with increasing $Q$; the mass function flattens towards extreme prolate axis ratios, while the concentration function remains steeper.  This suggests that when very high concentrations are measured for a given mass, a halo may be in this regime of significant elongation hidden along the line of sight.  A scaled likelihood value for each halo is also plotted ($0$ corresponds to the fit to a true spherical model, $1$ is the worst fit in our sample); oblate halos are fit quite poorly compared to their prolate counterparts due to their lower convergence and shear in the LoS orientation.

\subsection{Fits to Lensing Data from Halos with Varied Slopes}

\begin{figure}
 \includegraphics[width=80mm, height=80mm]{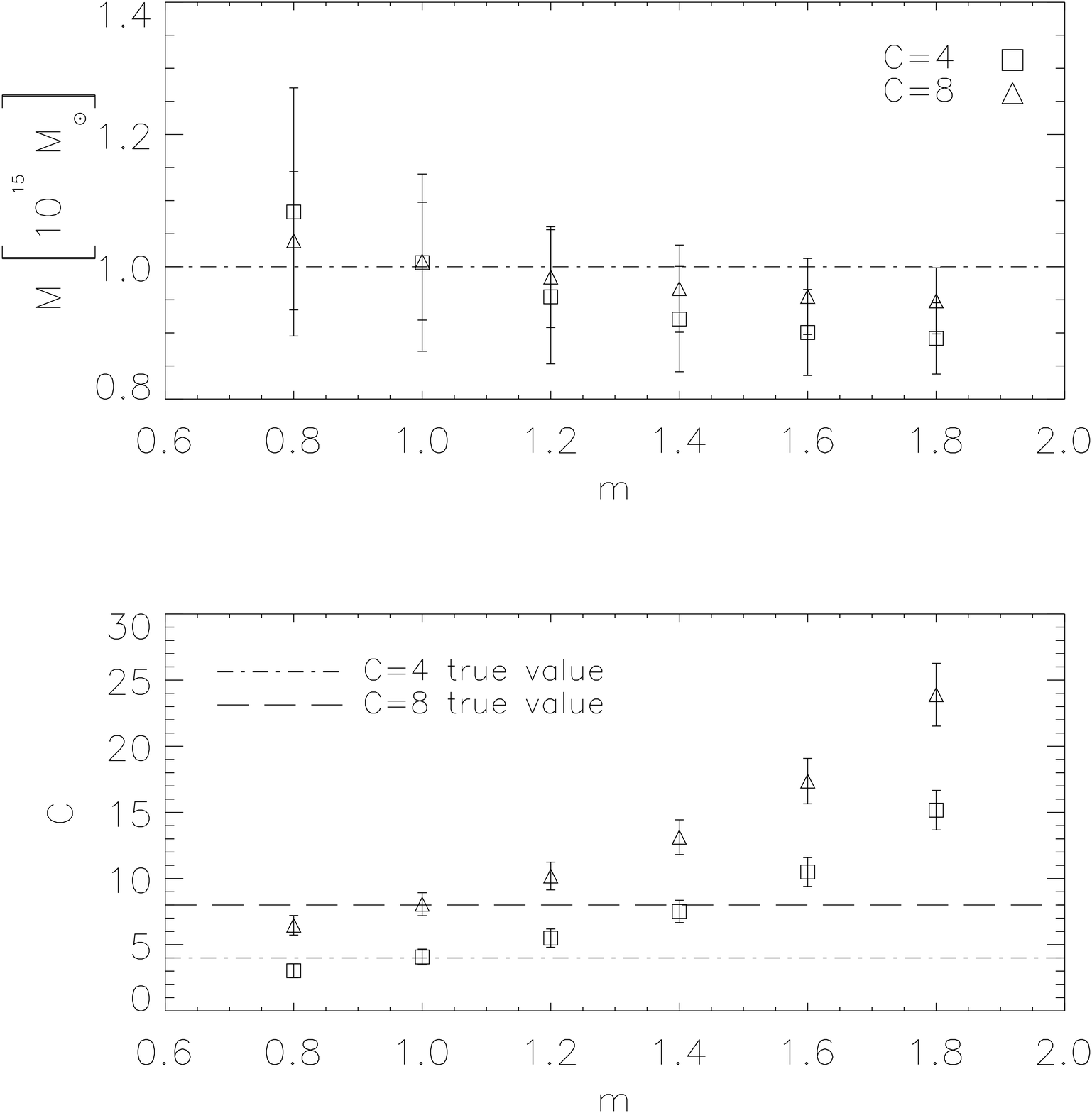}
 \caption{The top (bottom) panel plots the mean best-fit NFW mass $M_{200}$ (concentration $C$) after 500 lensing realizations through halos with variable inner slope, as a function of the inner slope $m$, for halo concentrations $C=4$ and $C=8$.  Error bars show the 1$\sigma$ dispersion within each set of realizations.}
 \label{fig:FreeSlopeNFW}
\end{figure}

\begin{figure}
 \includegraphics[width=80mm, height=35.5mm]{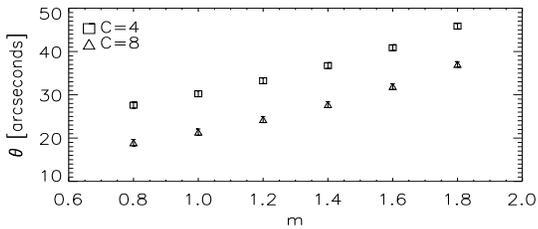}
 \caption{The mean best-fit SIS Einstein radius $\theta_E$ after 500 lensing realizations through halos with variable inner slope, as a function of $m$, for concentrations $C=4$ and $C=8$.  Error bars show the 1$\sigma$ dispersion within each set of realizations.}
 \label{fig:FreeSlopeSIS}
\end{figure}

\subsubsection{A steeper inner slope}
Six NFW-like halos with varied inner slope are studied with values of $m=\{0.8, 1.0, 1.2, 1.4, 1.6, 1.8\}$.  Though there are no predictions calling for inner slope values of less than one, $m=0.8$ is included to account for scatter about the mean and to allow observation of trends across the canonical NFW value $m=1$.  500 catalogues of background galaxies are lensed and NFW and SIS models are fit to the simulated lensed catalogues. We further repeat the simulations using halos of the same mass with a higher concentration $C=8$ to test how our results differ with changes in the underlying cluster parameters.

The best-fit NFW and SIS parameters are shown in Figures~\ref{fig:FreeSlopeNFW} and~\ref{fig:FreeSlopeSIS}.   We find that NFW-like halos with steeper inner slopes give rise to very high estimated concentrations (for $m=1.8$,  a cluster with a true $C=4$ is fit with a mean concentration $C_{fit} \approx 15$, and a halo with $C=8$ is fit with mean concentration $C_{fit}\approx 24$).   Thus assuming the paradigmatic value for the inner slope when in fact there is uncertainty about its mean value and significant scatter about that mean can have serious implications for the distribution of expected observed concentrations.  The mass is underestimated for steeper-sloped halos, and so therefore is the scale radius.  The SIS models exhibits the opposite trend, preferring larger $\theta_E$ values and masses for lenses with steeper inner slopes.  This likely occurs because the NFW-like halo is becoming more like an SIS as its inner slope approaches isothermal, which has a naturally higher Einstein radius at a given mass.

\subsubsection{Scatter in the outer slope}
Seven NFWs with varying outer slopes are simulated, each with $C=4$ and $M_{200}=10^{15}$ M$_{\odot}$, with values of $n=\{2.4, 2.6, 2.8, 3.0, 3.2, 3.4, 3.6\}$ to understand the effects of scatter about the canonical value on lensing analyses.  500 catalogues are generated and lensed as described above for each halo model and NFW and SIS models are fit to each.  Again, we repeat the simulations with a concentration of $C=8$.

\begin{figure}
 \includegraphics[width=80mm, height=80mm]{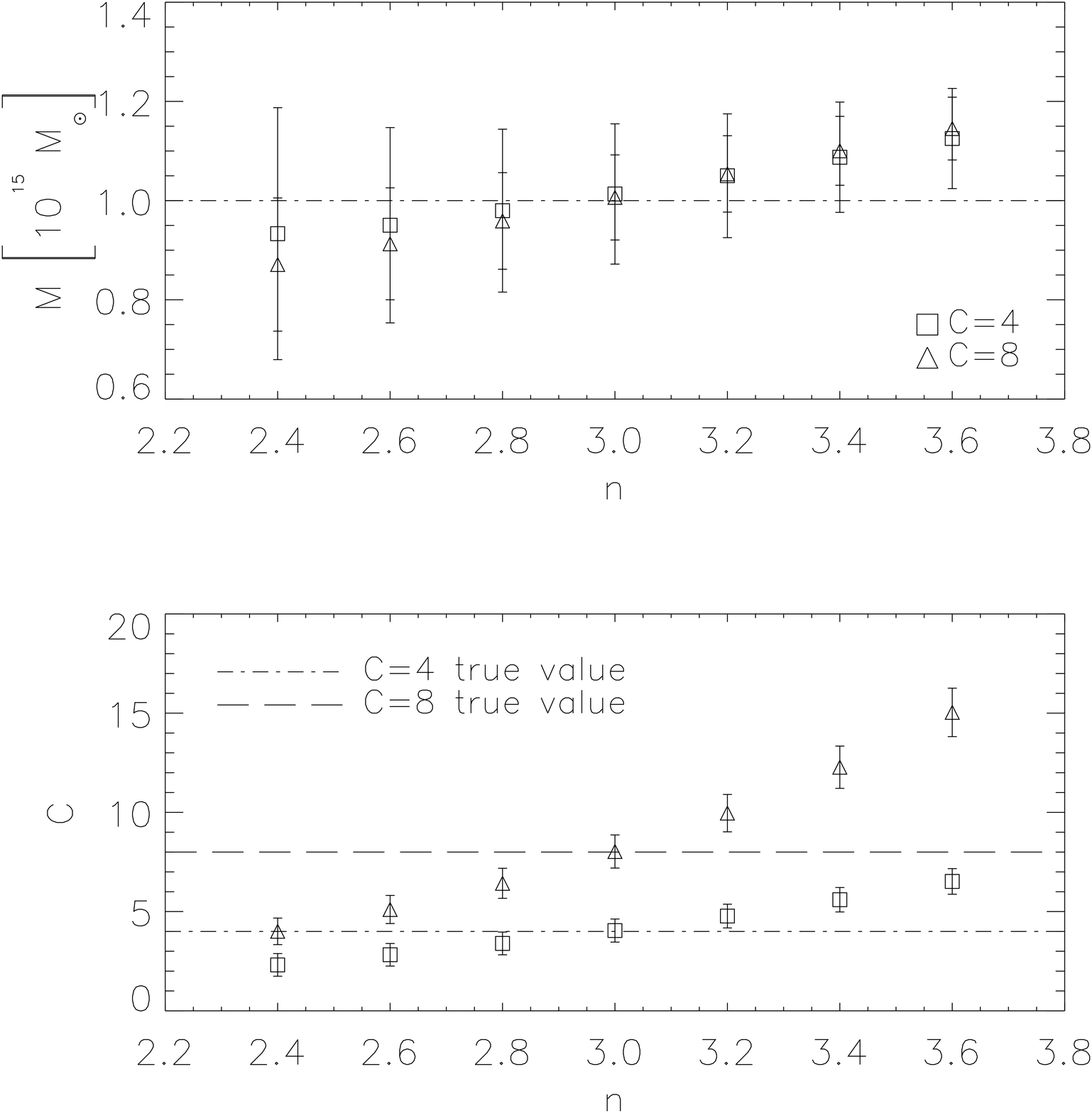}
 \caption{The top (bottom) panel plots the mean best-fit NFW mass $M_{200}$ (concentration $C$) after 500 lensing realizations through halos with variable outer slope, as a function of outer slope parameter $n$, for halo concentrations $C=4$ and $C=8$.  Error bars show the 1$\sigma$ dispersion within each set of realizations.}
 \label{fig:OFreeSlopeNFW}
\end{figure}

\begin{figure}
 \includegraphics[width=80mm, height=35.5mm]{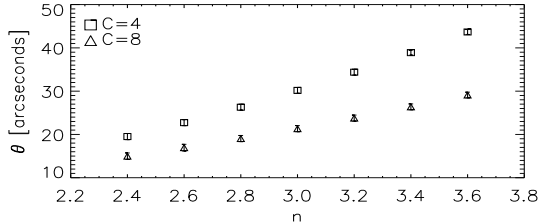}
 \caption{The mean best-fit SIS Einstein radius $\theta_E$ after 500 lensing realizations through halos with variable outer slope, as a function of outer slope parameter $n$, for halo concentrations $C=4$ and $C=8$.  Error bars show the 1$\sigma$ dispersion within each set of realizations.}
 \label{fig:OFreeSlopeSIS}
\end{figure}

The best-fit NFW and SIS parameters are shown in Figures~\ref{fig:OFreeSlopeNFW} and~\ref{fig:OFreeSlopeSIS}.  The NFW concentration is overestimated for steeper outer slopes and underestimated for shallower slopes; the mass follows the concentration, unlike in the case of a steeper inner slope.  The mass is a nearly linear function of steepness; an increase in steepness leads to a mass increase comparable to the decrease in $M_{200}$ caused by a slope decrease of the same magnitude.  The behaviour of the concentration, however, is not linear, with increases in steepness producing larger changes in $C$ than comparable decreases.  This means that even slopes normally distributed around the paradigmatic value $n=3$ will lead on average to measurements of slightly higher concentrations.  The SIS fits display the same mass trend, with mass and $\theta_E$ increasing with increasing steepness.  

\section{Discussion \& Conclusions}
We focus our further discussions on the impacts of triaxiality, the most potentially insidious of the three deviations from NFW we study, as it cannot be fully investigated and marginalized by any lensing technique due to degeneracies along the line of sight.

\begin{figure}
 \includegraphics[width=90mm, height=45mm]{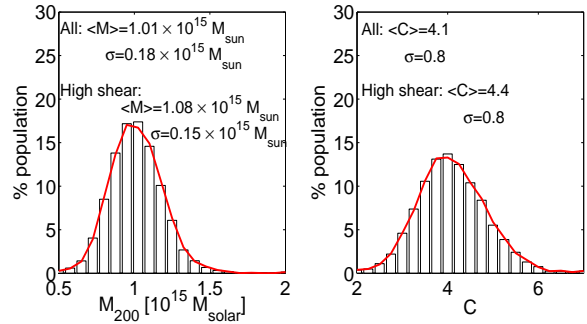}
 \caption{The distribution of best-fit NFW concentration and virial mass for 10,000 lensing realizations through $10^{15}$ M$_{\odot}$ halos with shapes drawn from the simulation-predicted triaxiality distribution and randomly oriented.  The solid line plots the same population, this time including only realizations for efficiently lensing halos with $<g>$ $\geq$ 0.014, the average shear for a $10^{15}$ M$_{\odot}$ spherical halo.}
 \label{fig:monte}
\end{figure}

To begin, it is crucial to know how much triaxiality is expected in the cluster-sized halos.  Recent N-body studies by \cite{s3} have resulted in detailed predictions of the axis ratio distribution for dark matter halos, finding that halos tend to be more prolate than oblate, with the distribution for $b/c$ peaking near $0.8$ and that for $a/c$ peaking near $0.65$ with a tail down to $0.4$.  While the inclusion of baryon interactions might reduce the levels of predicted triaxiality (H. Hoekstra, private communication), Shaw's predictions are a very useful limiting case in which to apply our results.  The top panels of Figure~\ref{fig:monte} plot the parameter distributions obtained from performing 10,000 lensing realizations through a $10^{15}$ M$_{\odot}$, $C=4$ halo, at each realization choosing axis ratios from the predicted distributions and randomly orienting the halo.  The means are equal to the true values, and the distributions are close to Gaussian, but with very long tails toward high values.  Though those tails are indeed extended by triaxiality, indicated by small increases in the dispersion of the distributions from those for distributions obtained by lensing through spherical halos, the total effect of triaxiality on a complete population of clusters is very small.

\subsection{Lensing Efficiency Effects}
Our finding that the mean best-fit $C$ and $M_{200}$ approach the true value for even very triaxial halos when averaged over orientation initially suggests that while triaxiality may increase the errors on parameter estimates for individual clusters, it should not be a significant issue in constraining the mass function when tabulated over a large sample.  However, this conclusion neglects to take into account the effects of orientation on lensing efficiency.

\begin{figure}
 \includegraphics[width=80mm, height=80mm]{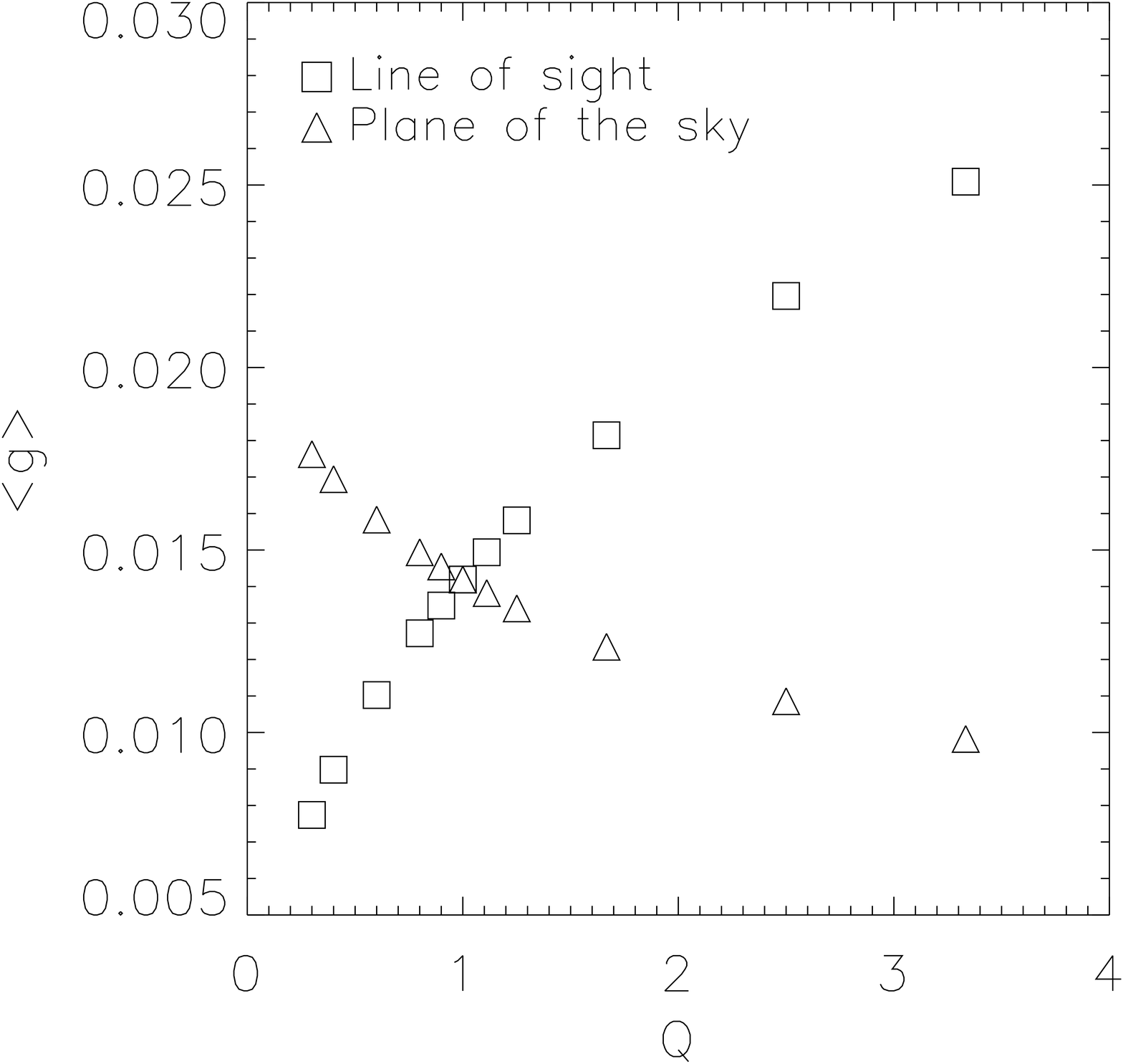}
 \caption{Average reduced shear $g$ within the annulus located between 1' - 7.5' of the halo center, the annulus from which the weak lensing data is drawn in the simulations, plotted as a function of axis ratio $Q$ for LoS and Plane halos.}
 \label{fig:lenseffic}
\end{figure}

Certain orientations of clusters will produce higher lensing efficiencies than others, making them more likely to be included in lensing-selected cluster surveys. Figure~\ref{fig:lenseffic} plots the average reduced shear over the annulus from which the weak lensing data is taken as a measure of lensing efficiency.

Halos with significant mass along the line of sight are the best lenses; prolate halos with high $Q$ (low $a$ and $b$) oriented along the line of sight have almost twice the average shear than does a spherical halo of the same mass.  Oblate halos with very low $Q$ (low $a$ and $b$) oriented in the plane of sky are also effective lenses, which can be understood by recalling that the long axis visible on the sky is also the length of the axis hidden in projection, giving rise to significant convergence and shear.

Halos that are the most efficient lenses will be more likely to be studied as lenses; thus clusters selected for their lensing properties are likely to reside in the high-triaxiality, low probability tails of the halo distribution.
While triaxial halos averaged over orientation in simulations give mean parameter values equal to the true values, this type of simulation does not recreate observational conditions in which prolate halos are much more likely to be observed as lenses when oriented near the line of sight, and oblate halos when oriented in the plane of the sky.  Furthermore, since very prolate and oblate halos (aligned advantageously) are much better lenses than spherical halos, observed halos, especially those with strong lensing effects such as Abell 1689, are likely to be represent the high triaxiality end of the halo distribution.

Understanding this behavior is important for survey design (see e.g. \cite{w2} for more on shear-selected surveys); any criteria involving lensing efficiency will skew both mass and concentration estimates high!  As an example, the bottom panels of Figure~\ref{fig:monte} shows distributions of NFW parameters for the same population of 10,000 lensing realizations shown above, this time keeping only parameters from realizations in which the lensing halos have more than a cutoff level of lensing efficiency, set to be the lensing efficiency of a spherical halo of $10^{15}$ M$_{\odot}$, $<g> = 0.014$.  Though even after this cut the effects of triaxiality are very small, the mean parameter values do indeed shift slightly towards higher mass and concentration.

\subsection{Model Discrimination}
A potential impact of the deviations from NFW that we study here that we have not yet addressed is that of changing the frequency with which one model family is incorrectly preferenced over another.  We find that varying the outer slope has no significant impact, and that triaxiality has only a very small impact, with very triaxial halos of $a=0.4$ being misidentified as SIS halos in 4\% more realizations than are spherical halos.  However, changes in the inner slope prove to be very significant; for an inner slope of $m=1.6$ NFWs are misidentified as SIS halos in 83\% more realizations than are canonical NFW halos.  Interestingly, an NFW with $m=1.8$ fares significantly \textit{better}, with misidentification in only 29\% more realizations.  It is not surprising that as the inner slope of the NFW model approaches the isothermal slope it would be confused more often with an SIS, and that fact that it fares worst when $m=1.6$ suggests that the SIS fitting routine is ``averaging'' the slope in some sense over the range in which the data is taken.  It is thus clear that weak lensing analyses are quite sensitive to the inner slope, even though the data are taken from outside the inner regions of the cluster, due to the non-local nature of the gravitational potential.

\subsection{The Meaning of Concentration}
Our choice to use the traditional $C=r_{200}/r_{s}$, as did \cite{w1}, rather than $C_{-2}$ of \cite{k2} in our treatment of an NFW with a free inner slope derives from two factors.  The first is that $C$ is the more commonly used parameter and thus more useful for comparisons; the second is that while \cite{k2} argue that the point at which the logarithmic slope is $-2$ is a more physical quantity than the scale radius $r_s$, we feel that the break radius at which the slope of the NFW is the average of the asymptotic inner and outer slopes is an equally valid physical scale to use.
  In any case, the numerical values of concentration are most important in comparison to those from simulations and between different members of the cluster population, using the same definition.

\subsection{Summary and Outlook}
In this paper we concentrate on constraints from weak lensing; much work remains to be done, marginalising over the effects of triaxiality for cluster data sets, pioneered by \cite{o2} for Abell 1689. In the past couple of years, a great deal of progress has been made in techniques combining weak and strong gravitational lensing constraints on cluster profiles (e.g. \cite{c4}; \cite{d3}). In addition, gravitational flexion has very recently been harnassed by \cite{l3} to probe substructure in clusters, and is also a promising means of constraining the mass distribution in the regime where standard weak lensing approximations break down. Besides the various regimes of gravitational lensing, large spectroscopic surveys have the potential to provide information essential to constructing models of the 3-D distribution of cluster mass and its dynamical history. So far this has only been undertaken for a very small number of clusters (e.g. \cite{c2}), but the advent of wide-field spectroscopy makes this feasible.  Such spectroscopic studies combined with deep space-based observations (where the number density of galaxies useful for shear measurements is several times that of ground-based data) and SZ and X-ray studies will allow for more detailed modeling of the mass distributions of clusters.

We investigate the effects of triaxiality and variations in the slopes of the NFW profile on weak lensing parameter and mass estimates and find that for individual clusters the effects can be significant.  Triaxiality causes the mass and concentration to be over or under estimated by $\sim50\%$ and a factor of 2, respectively; steepening the inner slope leads to up to $10\%$ underestimation of the mass and overestimation of the concentration by a factor of up to 3; scatter in the outer slope causes up to $10\%$ and $5\%$ errors in mass and concentration estimates, respectively.  Averaged over orientation and over predicted axis ratio distributions the errors induced by triaxiality are significantly reduced, which bodes well for the use of cluster masses as a cosmological probe; however, when lensing efficiency is taken into account they become more important.  Thus the selection effects must be accounted for in survey design, for example when shear selected samples are employed or in future follow-up of clusters detected using the S-Z effect.  Crucially, the effects of triaxiality along the line of sight and variations in halo slopes may lessen tensions between the unusual galaxy cluster concentration/mass estimates reported in recent work and the predictions of $\Lambda$CDM.

\section*{Acknowledgments}
This work was supported by the Marshall Foundation (VLC) and the Royal Society (LJK).  We thank Matthias Bartelmann, Vincent Eke, Chuck Keeton, Antony Lewis and  Damien Quinn for helpful discussions. We thank the referee for a helpful and constructive report.

\appendix

\section{Triaxial Parameter conversions}
We here compare our triaxial NFW parameters to those of \cite{j1} as well as parameters fit to spherically averaged 3D ellipsoidal density profiles, as is often done in fits to clusters in N-body simulations.

\begin{figure*}
 \includegraphics[width=160mm, height=70mm]{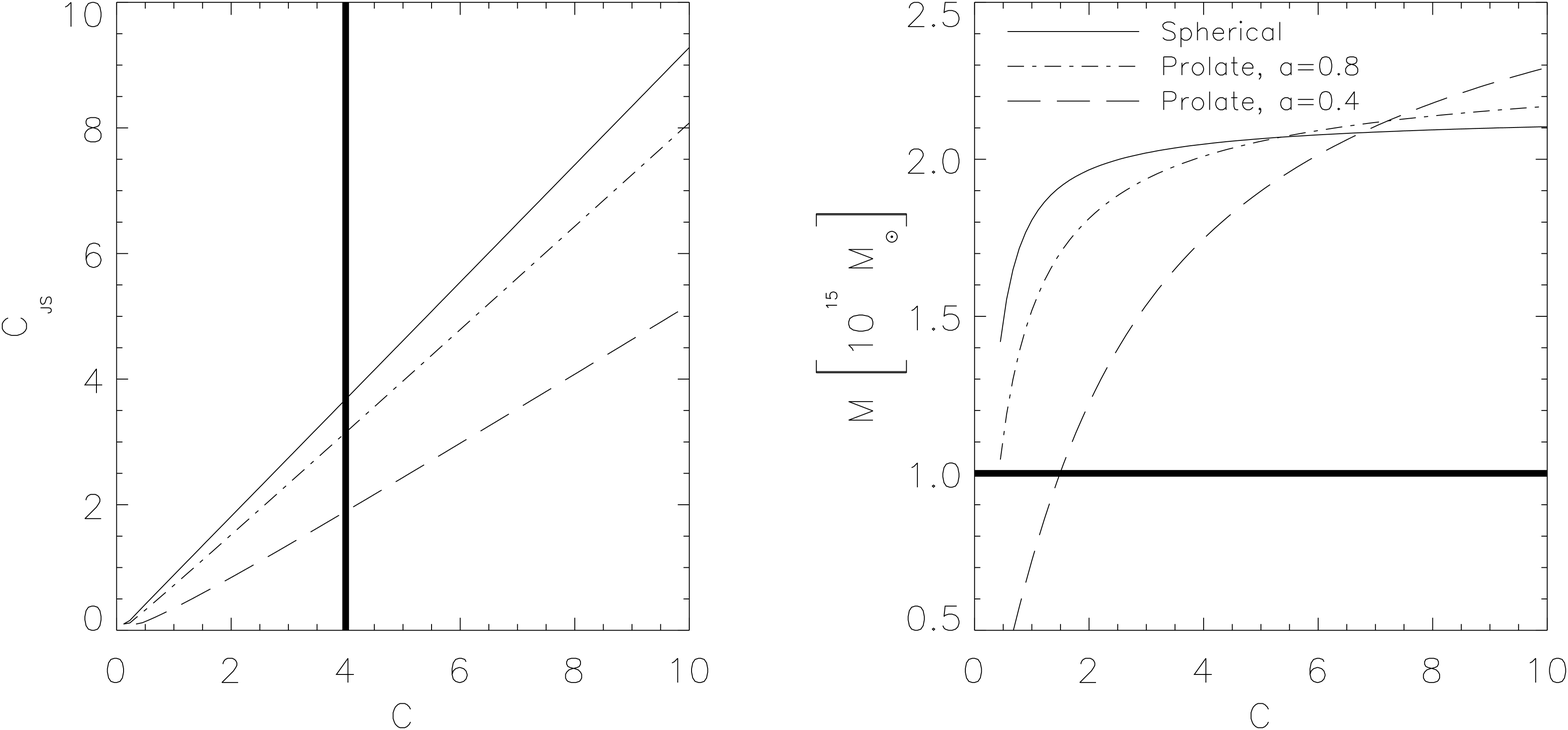}
 \caption{The JS02 concentration parameter $C_{JS}$ and virial mass $M_{JS}$ plotted as a function of $C$ for halos of $M_{200} = 10^{15}$ M$_{\odot}$.  Three examples are shown; one a very prolate halo with $a=b=0.4$, one a slightly prolate halo with $a=b=0.8$, and a spherical halo.  In the JS02 parameterization the halo is characterized as less concentrated and more massive for all triaxial models.}
 \label{fig:JS}
\end{figure*}
\begin{figure*}
 \includegraphics[width=160mm, height=70mm]{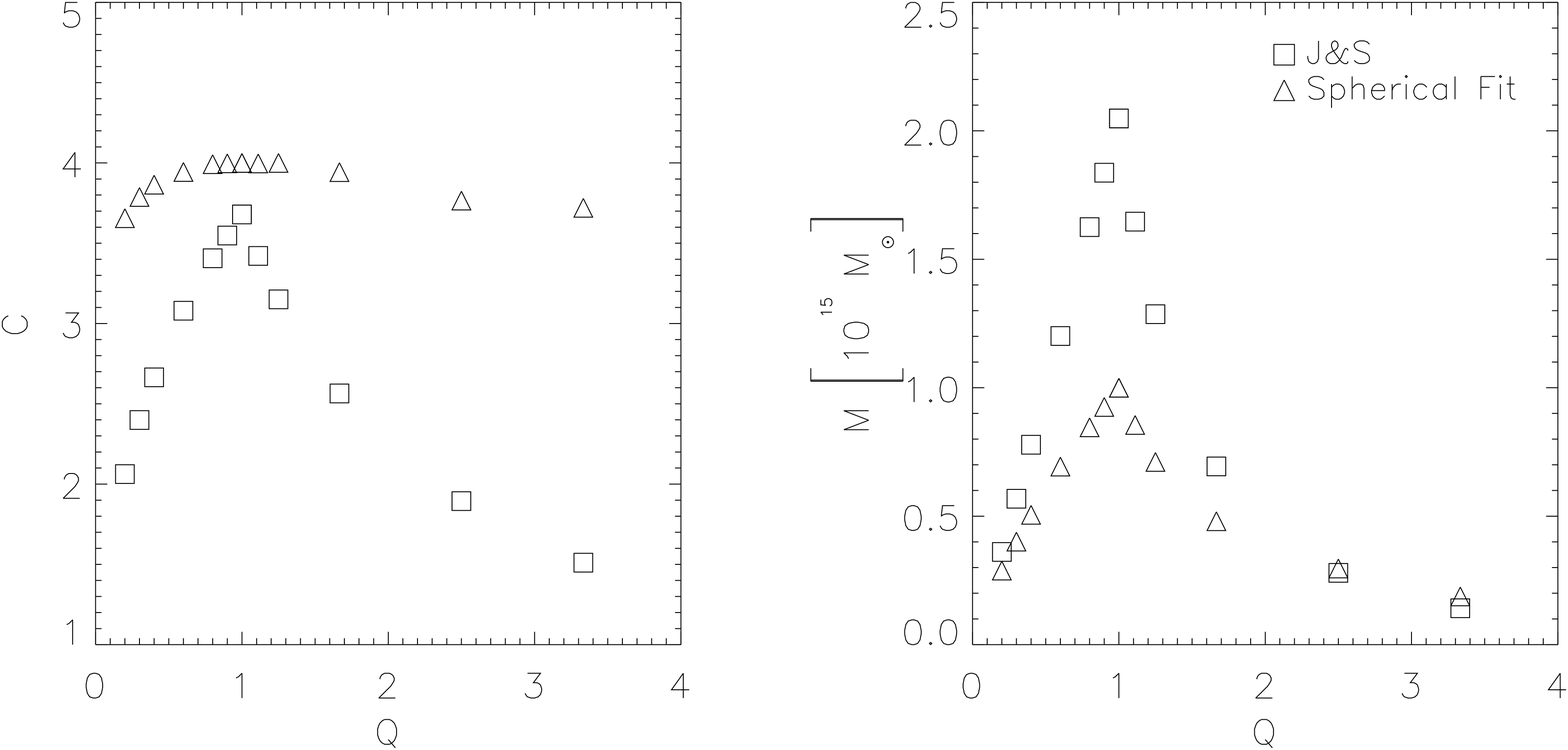}
 \caption{The spherical concentration parameter $C_{sphere}$ and virial mass $M_{sphere}$ as a function of triaxial axis ratio $Q$ for clusters of $M_{200} = 10^{15}$ M$_{\odot}$ and $C=4$.  The JS02 concentration and mass are also shown for comparison.}
 \label{fig:sphereNFW}
\end{figure*}
In Figure~\ref{fig:JS} the JS02 concentration parameter $C_{JS}$ and virial mass $M_{JS}$  are plotted against $C$ for a cluster of $M_{200}=10^{15}$ M$_{\odot}$ for two prolate clusters and a spherical NFW. As expected, there is close agreement between the parameters when the clusters are nearly spherical, but there are significant differences at more extreme axis ratios. 
To transform between parameter systems we require that a halo described in our coordinate system by $\Pi = \{C, R_s, R_{200}, M_{200}, a, b\}$ give the same density profile as a model described by $\Pi' = \{C_{JS}, R_0, R_{e}, M_{JS}, a, b\}$ in JS02's parameterization.  $C_{JS}$ is defined as
\begin{equation}C_{JS} \equiv \frac{R_e}{R_0}\end{equation}
where $R_e$ is defined such that the mean density within an ellipsoid of semi-major axis $R_e$ is
\begin{equation}\bar{\rho} = 5\Delta _{vir}\left(\frac{c^2}{ab}\right)^{0.75}\Omega(z)\rho_c(z)\end{equation}
where $\Delta_{vir}$ is the overdensity at virialization of the halo.  The characteristic overdensity of the halo, $\delta_{ce}$ is then given by
\begin{equation}\delta_{ce} = \frac{5\Delta _{vir}\left(\frac{c^2}{ab}\right)^{0.75}\Omega(z)}{3}\left(\frac{C_{JS}^3}{\ln(1+C_{JS}) - \frac{C_{JS}}{1+C_{JS}}}\right).\end{equation}
The second term is the same as for our parameterization, while the prefactor depends on the axis ratios unlike in our model.  Since $R_0=R_s$ is the same in both systems, setting the density profiles equal simply requires that $\delta_c$ = $\delta_{ce}$ for a given set of parameters.  Thus, converting from our parameterization to that of JS02 requires the numerical solution of
\begin{eqnarray}\frac{C_{JS}^3}{\ln(1+C_{JS}) - \frac{C_{JS}}{1+C_{JS}}} &=& \left(\frac{200}{5\Delta _{vir}\left(\frac{c^2}{ab}\right)^{0.75}\Omega(z)}\right)\times \nonumber \\ 
&&\left(\frac{C^3}{\ln(1+C) - \frac{C}{1+C}}\right).\end{eqnarray}
$R_e$ is then simply calculated, $R_e = C_{JS}R_0$, and $M_{JS}$ is given by the fitting formulae derived from JS02's numerical simulations:
\begin{equation}M_{JS} = \frac{4\pi \Delta_{vir} \Omega(z) \rho_c(z)}{3}\left(\frac{R_e}{0.45}\right)^3.\end{equation}

Figure~\ref{fig:sphereNFW} shows the best-fit spherical concentration parameter $C_{sphere}$ and virial mass $M_{sphere}$ as functions of axis ratio $Q$, defined in Section~\ref{sssec:LoSHalos}.  Parameters are calculated by averaging the density of the ellipsoidal cluster over spherical shells and fitting a spherical NFW density profile to the averaged values.  This is how fitting is often done to N-body simulations, in which parameter distributions are tabulated assuming spherically symmetric halos due to the significant degeneracies encountered in fitting triaxial density models (\cite{d1}).  The spherical concentration is very similar to our choice of $C$ at most axis ratios, but the spherical mass is significantly less than our choice of ellipsoidal $M_{200}$ for very large and small $Q$.

\bsp

\label{lastpage}

\end{document}